\documentclass{article}
\usepackage[utf8]{inputenc}
\usepackage[english]{babel}
\usepackage{csquotes}
\usepackage{authblk} % for author affiliations
\usepackage[%
  backend=biber,
  style=apa,
  natbib=true
]{biblatex}
\usepackage{hyperref}
\hypersetup{colorlinks=true,
	linkcolor=cyan,
	citecolor=blue,
	urlcolor=green}

\usepackage[inline]{enumitem}

\addto\extrasenglish{%
}	
\addto\extrasenglish{%
}	

\newcommand{\tokenaddress}[2]{\href{https://etherscan.io/token/#2}{\texttt{#1}}}

\usepackage{algorithm}
\usepackage{algorithmic}
	
\usepackage{tabularx}
\usepackage{booktabs}
\usepackage{multicol}
\usepackage{graphicx}

\title{From banks to DeFi: the evolution of the lending market}

\author[1]{Jiahua Xu}
\author[1]{Nikhil Vadgama}

\affil[1]{University College London, Centre for Blockchain Technologies}

\DeclareSourcemap{
  \maps[datatype=bibtex, overwrite=true]{
    \map{
      \step[fieldsource=doi,
        match=\regexp{\{\\_\}},
        replace=\regexp{_}]
    }
  }
}

\begin{document}

\maketitle

\begin{abstract}
    The Internet of Value (IOV) with its distributed ledger technology (DLT) underpinning has created new forms of lending markets. As an integral part of the decentralised finance (DeFi) ecosystem, lending protocols are gaining tremendous traction, holding an aggregate liquidity supply of over \$40 billion at the time of writing. In this paper, we enumerate the challenges of traditional money markets led by banks and lending platforms, and present advantageous characteristics of DeFi lending protocols that might help resolve deep-rooted issues in the conventional lending environment.
    With the examples of Maker, Compound and Aave, we describe in detail the mechanism of DeFi lending protocols. We discuss the persisting reliance of DeFi lending on the traditional financial system, and conclude with the outlook of the lending market in the IOV era.
\end{abstract}

\section{Introduction}

Lending and credit have been an integral part of human society for thousands of years, with the first evidence of loans in human history taking place in Mesopotamia approximately 5000 years ago \citep{freas2018credit}. Since then, lending markets have evolved to assume many different forms including consumer lending, student loans, mortgages, corporate debt and government bonds.

Fundamentally, lending is intricately related to the concept of trust and the promise of repayment. The act of lending is to lend money, to earn interest, and to be paid back. The term ``credit" itself comes from the mid-16th century French, meaning to believe and trust.

Lending has now become one of the most important financial activities in society. It fuels economic growth and cultivates forward-looking commercial activities. The size of the world's debt markets as of 2019 was estimated to be more than \$255 trillion or nearly \$32,500 for each of the world's 7.7 billion people, and more than three times the world's annual output \citep{jones2019debt}.

In recent times, technology's impact on lending markets has shown significant progress in solving many problems in relatively inefficient markets. For example, artificial intelligence and alternative data are making breakthroughs for financial inclusion through alternative credit scoring. This paper focuses on how blockchain technology and its role in the Internet of Value are helping create more efficient lending markets. In particular, we address consumer, business and more recent crypto-asset lending, but the principles discussed can, of course, apply to the broader unsecured and secured lending markets  \citep{FCA2021}.

The rest of this paper is structured as follows: in \autoref{sec:lendingconventional} we review conventional lending in society, fuelled by commercial banks as well as new marketplaces and lending businesses; in \autoref{sec:lendingmarket} we raise challenges currently faced in lending markets; in \autoref{sec:paradigmshift} we elaborate on new lending protocols in the decentralised finance (DeFi) space,
% ; in \autoref{sec:deficharacateristics} we 
and
argue how blockchain technology can empower lending markets for the age of the Internet of Value; in \autoref{sec:coevolution} we discuss DeFi lending protocols' status quo and their coevolution with centralised financial (CeFi) platforms; in \autoref{sec:outlook} we conclude with an examination on how lending markets may evolve and transform in the future.

\section{Lending in the conventional financial market}
\label{sec:lendingconventional}

Lending requirements of consumers and businesses are primarily met by two major types of organisation:
    \begin{enumerate*}[label={(\roman*)}]
    \item banks and
    \item new lending marketplaces and specialised companies.
    \end{enumerate*}
In this section, we discuss the principal lending mechanics of these organisations.
\subsection{Banks}

In the lending market, banks hold a unique position as that of a ``money creator'' \citep{Werner2014a}. When providing a loan, a bank simultaneously extends its balance sheet by increasing its assets (loans receivable) and its liability (borrower's deposit) \citep{Lindner2013}. The amount that commercial banks can lend is not as simple as a multiplier effect of the reserves they hold at the central bank; instead, it depends on various factors to do with market forces, risk, interest rates, borrower behaviour and regulatory policy. The majority of broad money in economies (with particular reference to the UK) are bank deposits, accounting for 97\% of the total amount of money in circulation, created via loans \citep{Mcleay2014moneyCreation}. Let us look at Central banks and Commercial banks in more detail below.

\paragraph{Central banks}
A new loan issued by a central bank can increase the central bank's outstanding money supply. Thus, a central bank is the ``lender of last resort" as it is technically capable of issuing large-scale loans commonly used as a rescue plan to ``bail out" commercial banks in a financial crisis. This practice has been occasionally employed since the global financial crisis of 2007-2009 in the form of Quantitative Easing (QE), where central bank reserves are created to purchase financial assets, mainly from non-bank financial companies \citep{Mcleay2014moneyCreation}. Naturally, a central bank's excessive loan issuance inflates monetary supply, which potentially leads to currency depreciation. Bailout plans of this sort--where central banks ``print" fresh money to ensure market liquidity--have historically received criticism as the action dilutes the money's value.

\paragraph{Commercial banks}
For commercial banks, newly created deposits through loan issuance can also be used as money in the sense that they can be accepted as a form of payment by non-banks. In effect, a deposit at a commercial bank represents the bank's promise to pay central bank money. Failure to collect a sufficient amount of outstanding loans as a result of e.g. borrowers not being able to repay,  thus induces a breach of promise on the part of the commercial bank, potentially resulting in bankruptcy and, more severely, mistrust in the overall banking system. This was precisely the story of the global financial crisis, where banks collapsed due to the high rate of default in the secondary mortgage market.

In summary, loans from banks (central or commercial) generate deposits, fundamentally different from the deposit-first operation of lending companies, which we discuss below.

\subsection{Lending companies and marketplaces}

Distinct from banks, ordinary lending companies function only as an intermediary of loanable funds; they cannot create forms of money.  In other words, a lending company must absorb liquidity from lenders first in order to provide its borrowing clients' loans.

Indeed, it is mechanically feasible for lending companies to also operate as commercial banks:  issuing a loan by creating a new deposit in the form of ``promise to  pay  central  bank  money”,  in which case deposits would not need to fully cover loans. Critically, however, the ``promise" from lending companies—unlike commercial banks—can not be used as legal tender. Consequently, borrowers would find it difficult to use this newly created deposit as a form of payment anywhere outside the lending platform. As \cite{Minsky} put it: ``everyone can make money; the problem is to get it accepted". Nevertheless, there are many other derivatives such as collateralised debt obligations (CDOs) that can be fungible in capital markets. It is also possible to transfer plain vanilla loans to those who might accept it as a substitute for fiat funds.

Generally speaking, there exist three mechanisms within non-bank lending.

\paragraph{Quote driven market}

In a quote-driven market, the lending platform attracts both lenders and borrowers at the same time. Lenders on these types of platforms can be either institutional wealth managers or retail investors, as seen on UK-based \href{https://www.fundingcircle.com/global/capitalmarkets/}{Funding Circle} and US-based \href{https://help.lendingclub.com/hc/en-us/articles/115009000328-How-LendingClub-balances-different-investors-on-its-platform}{LendingClub}. The market trend has been that more and more institutional funding is being drawn to these platforms, and preferred by the platforms due to the economies of scale when dealing with larger investors \citep{ziegler2020alternativeFinance}. The borrowers are often individuals or small and medium enterprises, who apply for a loan by supplying relevant information such as loan principal required, borrowing period and credit history. While acting as an agent without taking any credit risk, a lending platform usually performs some degree of risk assessment of borrowers based on their profile to ensure the minimum quality of listed loans.\footnote{\url{https://www.fundingcircle.com/uk/resources/investors/team/meet-the-credit-assessment-team/}} Rather than linking individual quotes from lenders and borrowers, most platforms nowadays pull funds and manage a diverse set of borrowers to better diversify risks for their pool of investors. Loans are fractionalised such that lenders can invest in a large number of these smaller loan chunks to reduce idiosyncratic risks, and can achieve exposure based on their risk appetite.

\paragraph{Order driven exchange}
In an order-driven exchange platform, orders to borrow and lend with various price levels (in the form of e.g. interest rate per annum) are cleared automatically through an order book. To address default risk, borrowers often need to provide sufficient collateral before placing an order. This mechanism is typically adopted for trading in global debt markets.

\paragraph{Over-the-counter}
An over-the-counter model refers to a purely bilateral lending model. Similar to how quote driven markets operate, firstly a borrower submits a request to the platform enclosing personal information, credit history, etc. The platform then evaluates the request and assigns an interest rate suitable for the borrower's particular request and credibility. The request is subsequently posted on the platform where lenders can choose to fulfil individual requests. Peer-to-peer lending platforms, such as the China-based \href{https://invest.ppdai.com/loan/listpage}{PPDai}, adopt this model. Earlier lending platforms typically engaged in this type of model before moving to a quote-driven market model.

\section{Challenges with the mainstream lending market}
\label{sec:lendingmarket}

The current mainstream lending market, led by banking institutions, is fraught with issues.

\paragraph{Financial inclusion}

After the global financial crisis, more stringent lending rules and underwriting models were applied to banks. As a consequence, the threshold of credit rating was increased to lower the default risk. Clients with thin credit files, such as small and medium enterprises (SME) and retail borrowers (particularly immigrants and women) have become the most negatively impacted, facing increased difficulty to borrow from banks. The particular role that SMEs play in most major economies, especially developing economies, cannot be underestimated. Unfortunately, they are significantly affected by financial inclusion issues in lending, and lack access to alternative lending channels such as international capital markets. SMEs represent 90\% of businesses and more than 50\% of employment worldwide. The International Finance Corporation estimates that 65 million firms have unmet financing needs of \$5.2 trillion each year.\footnote{\url{https://www.worldbank.org/en/topic/smefinance}}

\paragraph{Passthrough issues}

Owing to the inefficiency of information flow and the tardiness in policy execution, government lending schemes often do not reach end clients fast enough. 
This was particularly evident during the Coronavirus pandemic where commercial banks responsible for distributing loans to SMEs in the UK have been not been efficiently passing them along \citep{Barrett2020}. 
The passthrough of changes in the interest rates were also delayed, resulting in a high lending rate in a low savings interest environment, which is detrimental to borrowers. For example, in the UK, the Bank of England’s rate cuts are often not passed on to lower mortgage costs for borrowers \citep{manas2020bankOfEngland}.

\paragraph{Intermediary cost}

Regulatory obstacles keep the barrier of entry for lending entities high, leading to an oligopoly market with imperfect competition. The centralisation of lending services enables high intermediary costs, and the resultant market friction further contributes to the inefficient usage of market liquidity, leading to the non-maximisation of utilitarian societal welfare. For example, in 2016, the UK government set out to challenge high street banks' oligopoly by encouraging the alternative lending sector's growth. To date, these new players have only managed to capture 20\% of the market \citep{prill2020openBanking}.

\paragraph{Liquidity inefficiency}

The existing lending mechanisms in the mainstream lending market produces sub-optimal liquidity outcomes. Both the supply and demand sides of liquidity are crudely diverged into siloed submarkets based on factors such as the lending period, interest rate, credit rating, etc., even within the same lending platform. The oversupply of liquidity in one submarket cannot be promptly transferred to serve the demand from another submarket.

\paragraph{Subprime problems}

The challenges in securing a loan with a bank or any other major financial institutions leave many potential borrowers unserved. This prompts the rise of alternative lending entities, including peer-to-peer lending markets, that cater to the cohort unable to obtain financing from banks. Those lending entities typically charge borrowers a premium for securing funding, understanding that the borrowers have already been rejected by other sources and are left with no other options. Unfortunately, fraudulent activities and high default rates permeate these less strict lending markets, stifling their growth.

\paragraph{Legacy infrastructure}

The dated information technology infrastructure used by mainstream lending entities is a crucial impediment to efficiency gains. Due to a lack of data exchange between financial institutions and the absence of a functional tracking system, clients' credit history is fragmented and opaque, exacerbating financial exclusion and fraud. Even with the entrance of new lending platforms with substantially lower operating costs (at levels half that of commercial banks), their ability to compete in this domain is outweighed by commercial banks' far cheaper funding costs \citep{deloitte2016marketplaceLending}.

\section{Paradigm shift in lending enabled by the Internet of Value}
\label{sec:paradigmshift}

New emerging technologies are improving the state of lending markets.  In particular,  blockchain technologies underpin the Internet of Value by creating a new economic layer of value exchange on top of the Internet. The rapid development on the Internet of Value is shifting the lending paradigm as new lending schemes enabled by decentralised finance, or DeFi in short \citep{Werner2021b}, solve entrenched issues native to the current mainstream lending market. These new solutions present themselves in the form of protocols \citep{Gudgeon2020a}, sets of rules determining how a lending market operates.

The DeFi market picked up momentum in 2020 and as of January 2021, the DeFi market's size, measured by the Total Value Locked, was estimated to be approximately \$44 billion.\footnote{\url{https://defipulse.com}} DeFi projects include decentralised lending platforms, decentralised exchanges, derivatives, payments, and assets.

\subsection{Key components of DeFi lending protocols}
\label{sec:components}

Despite their respective particularities, most DeFi lending protocols share two features: 
    \begin{enumerate*}[label={(\roman*)}]
    \item they deviate from existing subjective frameworks of centralised credit assessment to codified collateral evaluation;
    \item they employ smart contracts to manage crypto-assets \citep{Bartoletti2020sokLendingPools}.
    \end{enumerate*}
A smart contract is a program deployed on a distributed ledger, commonly the Ethereum blockchain, that can perform bookkeeping and calculations, receive and hold digital assets, and execute transactions automatically when triggered by predefined events.

\paragraph{Value locked}
Value locked originates from a users' deposits in a protocol's smart contract(s). The locked value forms a reserve to pay back depositors in redemption, and can also be used as collateral.

\paragraph{IOU token}
Lending protocols issue users IOU tokens against their deposits. Only those IOU tokens can be used to redeem deposits at a later stage, and they are also transferable and usually tradeable in exchanges.   

\paragraph{Collateral}
A loan's collateral comprises the entirety or part of the borrower's deposit. The collateral value, together with the underlying asset's maximum loan-to-value ratio determines how much crypto-assets a user is allowed to borrow.

\paragraph{Liquidation}
The liquidatable status of collateral is triggered automatically by a smart contract. When a loan's loan-to-value ratio exceeds a critical threshold--typically referred to as ``liquidation threshold''--due to interest accrued or market movements, any network participant can compete to liquidate the collateral. The market price information of locked and borrowed assets are supplied to smart contracts through external data feeds providers called ``price oracles".

\paragraph{Interest rate}
Borrowing and lending interest rates are computed and adjusted by smart contracts according to the supply-borrow dynamics, based on protocol-specific interest rate models. 

\paragraph{Governance token}
Certain DeFi lending protocols distribute to users governance tokens that allow them to propose and vote on protocol changes, such as modification of interest rate models. Governance tokens are often used as a reward scheme to incentivise participation, from both borrowing and lending sides, in a protocol.

\autoref{tab:defilending} displays the largest decentralised lending protocols by funds locked.

\begin{table}[h]
    \centering
    \begin{tabularx}{\textwidth}{Xrrrr}
    \toprule
    Protocol            &   Value locked  &   IOU        & Governance                  & Market cap  
    \\
    &   (billion USD)  &    token       &  token                  & (billion USD)   
    \\ \midrule
    Maker               &   9.37              &   {\tt DAI}       & \tokenaddress{MKR}{0x9f8f72aa9304c8b593d555f12ef6589cc3a579a2}                                & 3.46              
    \\
    Compound            &   11.05              &   {\tt cToken}s   &
    \tokenaddress{COMP}{0xc00e94cb662c3520282e6f5717214004a7f26888}   
    & 5.77              
    \\
    Aave                &   6.41              &  {\tt aToken}s    &
    \tokenaddress{AAVE}{0x7fc66500c84a76ad7e9c93437bfc5ac33e2ddae9}        & 
    7.27             
    \\
    \bottomrule
    \end{tabularx}
    \caption{Overview of major DeFi lending protocols on Ethereum. Value locked refers to total balance of {\tt ETH} and ERC-20 tokens held in lending pool contracts and is retrieved from \url{https://defipulse.com/}. Market cap refers fully diluted market cap of the governance token and is retrieved from \url{https://etherscan.io/}. Data are updated on 17 April 2021.}
    \label{tab:defilending}
\end{table}

\subsection{Major DeFi lending protocols}

\subsubsection{Maker}
Maker protocol revolves around {\tt DAI}, a stablecoin whose value is soft-pegged to the US dollar. Users who wish to borrow {\tt DAI} must first lock collateral in a smart contract named the Maker collateral vault. The collateral can be {\tt ETH} (the native coin on Ethereum) or certain ERC20 tokens (digital assets on Ethereum). The collateral can constitute one single asset or multiple assets. The smart contract calculates the collateral value based on its quantity and its market price, where the market price is input through an oracle, i.e. an external data feed provider. The borrower can then request the smart contract to issue him some {\tt DAI} amount below a predetermined threshold fraction of the collateral value. This threshold fraction determines when the collateral will liquidate. Borrowers are thus advised to request the issuing amount of {\tt DAI} to be moderately below the threshold, such that the slight downward price movement of the collateral against {\tt DAI} will not render the collateral available for liquidation.

There is no margin call with the Marker protocol. It is the borrower's responsibility to keep the borrow position overcollateralised by topping up collateral or repaying {\tt DAI} in case of unfavourable price change. As soon as a borrower's loan-to-value ratio exceeds the liquidation threshold, any network participant is entitled to bid for the collateral by repaying part of the loan, thus liquidating the position.

To redeem the collateral from the vault, the borrower needs to repay their {\tt DAI} loan and the loan interest, termed the ``stability fee". The stability fee accrues over time. Its value is dynamically adjusted. When the stability fee is high, the borrower is incentivised to pay back some {\tt DAI}, which is subsequently burned by the smart contract. Thus, this stability fee serves as a mechanism to steer the circulating supply of {\tt DAI}, ensuring that the currency value does not deviate too much from its peg. There is no fixed loan period; {\tt DAI} can be repaid partially or entirely at any time.

\subsubsection{Compound}

Users of the Compound protocol can supply and borrow {\tt ETH}, as well as an array of ERC-20 tokens. Users who deposit crypto-assets into the protocol's smart contract will receive some {\tt cToken} (e.g. {\tt aETH}, {\tt cDAI}) of an equivalent value, which can be used in exchange for the deposited asset plus interest in the future. A {\tt cToken} is an interest-bearing token, whose exchange rate against the deposited asset increases over time \citep{Leshner2019}.

To borrow from the protocol, users first have to deposit funds, which are used as collateral for the borrow position. In that sense, a borrower must first and foremost also be a depositor. The Compound protocol computes and updates borrowing and lending interest rates for each asset automatically and continuously, based on the amount deposited and locked, as well as  the amount borrowed \citep{Perez2021}.

The funds borrowed thus accrue interest in a time-variant manner. There is a maximum amount of funds that the user can borrow against his collateral. The market value of assets is determined through the protocol's price oracle. As both borrowed assets and collateralised assets experience price movements and accrue interest, the collateralisation ratio of a borrower's position changes continuously. Despite the absence of a fixed loan term, the borrower needs to ensure the overcollateralisation of his borrow position, such that the collateral does not become available to be liquidated by other network participants.

\subsubsection{Aave}

With the Aave protocol, formerly ``ETHLend'', liquidity suppliers deposit funds in a smart contract and receive the corresponding {\tt aToken} (e.g. {\tt aETH}, {\tt aDAI}), representing a deposit certificate. An {\tt aToken} is an interest-bearing token whose value is one-to-one pegged to the deposited asset. For example, if a user deposits 12 {\tt ETH}, they will receive 12 {\tt aETH} as proof of deposit. The balance of {\tt aETH} automatically increases over time, mirroring the interest payment against the {\tt ETH} deposit \citep{Wowaave.com}. {\tt aToken} holders can always redeem the underlying asset with a 1:1 exchange rate by sending {\tt aToken} to the smart contract and receiving an equivalent amount of the underlying asset from the smart contract.

As with Compound, Aave users' borrow positions must be collateralised by their supplied funds and face liquidation risk once the position becomes undercollateralised. With Aave, borrowers can choose to use part of or the entirety of their deposited assets as collateral, and may switch between a variable interest rate and stable interest rate at any time.

Aave also supports ``flash loans''. \cite{Wolff} introduced flash loans as a type of loan where a user borrows and repays funds within one transaction without collateral. Flash loans are often used to take arbitrage opportunities
% e.g. across different trading pairs
% across decentralised exchanges 
(see {\bf Flash loan} \ref{algo:arbitrage}),
and to liquidate insufficiently collateralised borrow positions on lending protocols (see {\bf Flash loan} \ref{algo:liquidation}).
% , enabled through smart contracts. 

\begin{algorithm}[H]
\floatname{algorithm}{Flash loan}
\caption{Arbitrage}
\label{algo:arbitrage}
% {\it Example transaction:} % TODO
\begin{algorithmic}[1]
\STATE {\bf Borrow} crypto-asset {\tt XYZ}
\STATE {\bf Sell} {\tt XYZ} on exchange A at $P_A$
\STATE {\bf Buy} {\tt XYZ} on exchange B at $P_B$ with $P_B < P_A$
\STATE {\bf Repay} {\tt XYZ}
\end{algorithmic}
\end{algorithm}

\begin{algorithm}[H]
\floatname{algorithm}{Flash loan}
\caption{Liquidation}
\label{algo:liquidation}
\begin{algorithmic}[1]
\STATE {\bf Borrow} $x_1$ crypto-asset {\tt XYZ}
\STATE {\bf Liquidate} a borrow position by paying $x_1$ {\tt XYZ} to receive the collateral {\tt ABC} with a bonus
\STATE {\bf Swap} received {\tt ABC} for $x_2$ {\tt XYZ} with $x_2 > x_1$
\STATE {\bf Repay} $x_1$ {\tt XYZ}
\end{algorithmic}
\end{algorithm} 

% https://etherscan.io/tx/0x3baef5bccd9d1127591797ed5159ea14cd42539140309afe600e8fbf63ab6918

Users themselves are responsible for customising each specific flash loan contract for a determined cycle of operations. A flash loan transaction is atomic: if the loan is not paid back, the entire transaction is reverted, while transaction fees still incur.

\subsection{Current use cases}
\label{sec:application}

As of today, borrowers have been using DeFi lending protocols for two main purposes.

\subsubsection{Earning rewards}
At the moment, the primary motivation for ordinary retail users of DeFi lending protocols has been to receive participation rewards in the form of e.g.\ valuable, tradable governance tokens (see \autoref{sec:components}).

In the extreme, borrowers would repetitively 
\begin{enumerate*}[label={(\roman*)}]
\item borrow, 
\item re-deposit borrowed funds as collateral, 
\item borrow again.
\end{enumerate*}
As such, a ``borrow spiral'' is formed to maximize the amount of reward tokens that a user can receive \citep{Cousaert2021}.

\subsubsection{Leveraged trading}

Leveraged trading is commonly seen among more 
sophisticated investors as well as institutional investors such as hedge funds.
For example, an investor bullish about {\tt ETH} may borrow, say, {\tt DAI} to buy some {\tt ETH}. In expectation of a price increase of {\tt ETH}, investor would swap borrowed {\tt DAI} for  {\tt ETH} on an exchange, hoping that the purchased {\tt ETH} can be worth more {\tt DAI} in the future to such an extent that it exceeds the loan amount and leaves the investor some profit.

Similar to the borrow spiral discussed above, an investor can repetitively 
\begin{enumerate*}[label={(\roman*)}]
\item borrow {\tt DAI}, 
\item swap {\tt DAI} for {\tt ETH}, 
\item re-deposit {\tt ETH} as collateral,
\item borrow more {\tt DAI}.
\end{enumerate*} 
As such, a ``leveraging spiral'' is formed \citep{Perez2021} to maximise the investor's long exposure to a crypto-asset that is expected to appreciate.

% \begin{enumerate*}[label={(\roman*)}]
%     \item to earn governance tokens (see 
%     \item to perform speculative leverage trading.
% \end{enumerate*}

\subsection{Advantages of DeFi lending protocols}
\label{sec:deficharacateristics}

DeFi lending protocols exhibit apparent advantages over conventional lending schemes, broken down as follows.

\paragraph{Transparency} Deployed on a public blockchain, the exact content of smart contracts according to each DeFi lending protocol is freely available and auditable. In addition, users' historical interactions with protocols and their lent and borrow positions are transparently recorded on the blockchain. Market information is public for everybody.

\paragraph{Democracy} Absent a central authority, users vote on amendments of a protocol. In particular, governance tokens that are issued by many of these protocols to users give proportionate voting rights to those who have economic stakes in these platforms.

\paragraph{Liquidity} Funds supplied to a lending protocol are pooled together and can be utilised efficiently. Thanks to smart contracts and blockchain, lending, borrowing and arbitraging can all be performed inexpensively and almost instantaneously. By guaranteeing IOU tokens' redeemability to fund suppliers, DeFi lending protocols also ensure full transferability and exchangeability of debt holdings.

\paragraph{Agility} DeFi lending protocols can automatically update interest rates and ceaselessly reflect the latest supply-borrow balance of the market. 

\paragraph{Trustlessness} Lenders do not need to trust borrowers' solvency anymore, as ingeniously designed smart contracts enforce liquidation when default risk is present.

\section{Discussion}
\label{sec:coevolution}

\subsection{Status quo of DeFi lending protocols}
The pace of innovation of DeFi lending protocols is astonishing, with new projects emerging and rapid gaining attention, being copied and improved with new ideas quickly integrated into existing protocols. Nonetheless, the decentralised lending market is still in its infancy and at an experimental stage. 

In particular, 
the growth of today's DeFi lending space is still mainly propelled by short-term oriented, ``financialised'' borrowing (see \autoref{sec:application}). Nevertheless, the development of the entire DeFi ecosystem coupled with the increasing adoption of crypto-assets will encourage more dominated genuine, utility-driven borrowing activities with DeFi lending protocols.

Given that everybody must ``play by the code" in the decentralised environment, security requirement for protocols is exceptionally high. Unfortunately, there persist vulnerabilities and loopholes in smart contracts which attackers--rightfully albeit not righteously--exploit and profit from \citep{Qin2020b}. In particular, malicious users are seen to abuse certain features of DeFi lending protocols to perform e.g.\ flash-loans-funded price oracle attacks \citep{xu2021dexAmm}.  
The pseudonymous nature of blockchain, plus the absence of central authority, makes it challenging for attack victims to claim damage. As it stands, the community still resorts to regulations to deter malicious activities. In the long run, an improved protocol governance will lead to system resilience, and a dependence on law enforcement external to the decentralised ecosystem will be reduced.

\subsection{Coevolution of DeFi and CeFi}

DeFi still heavily relies on the long-established banking system. Notably, the value of crypto-assets on DeFi are still primarily gauged and acknowledged in fiat values. Among the most widely used crypto-assets are stablecoins (e.g. {\tt DAI}), whose value is anchored to fiat currencies. As discussed at the beginning of this article, only central banks can, by definition, issue central bank money. DeFi's reliance on fiat denomination makes central banks irreplaceable, at least in the near future.

\begin{figure}
    \centering
    \includegraphics[width=\linewidth]{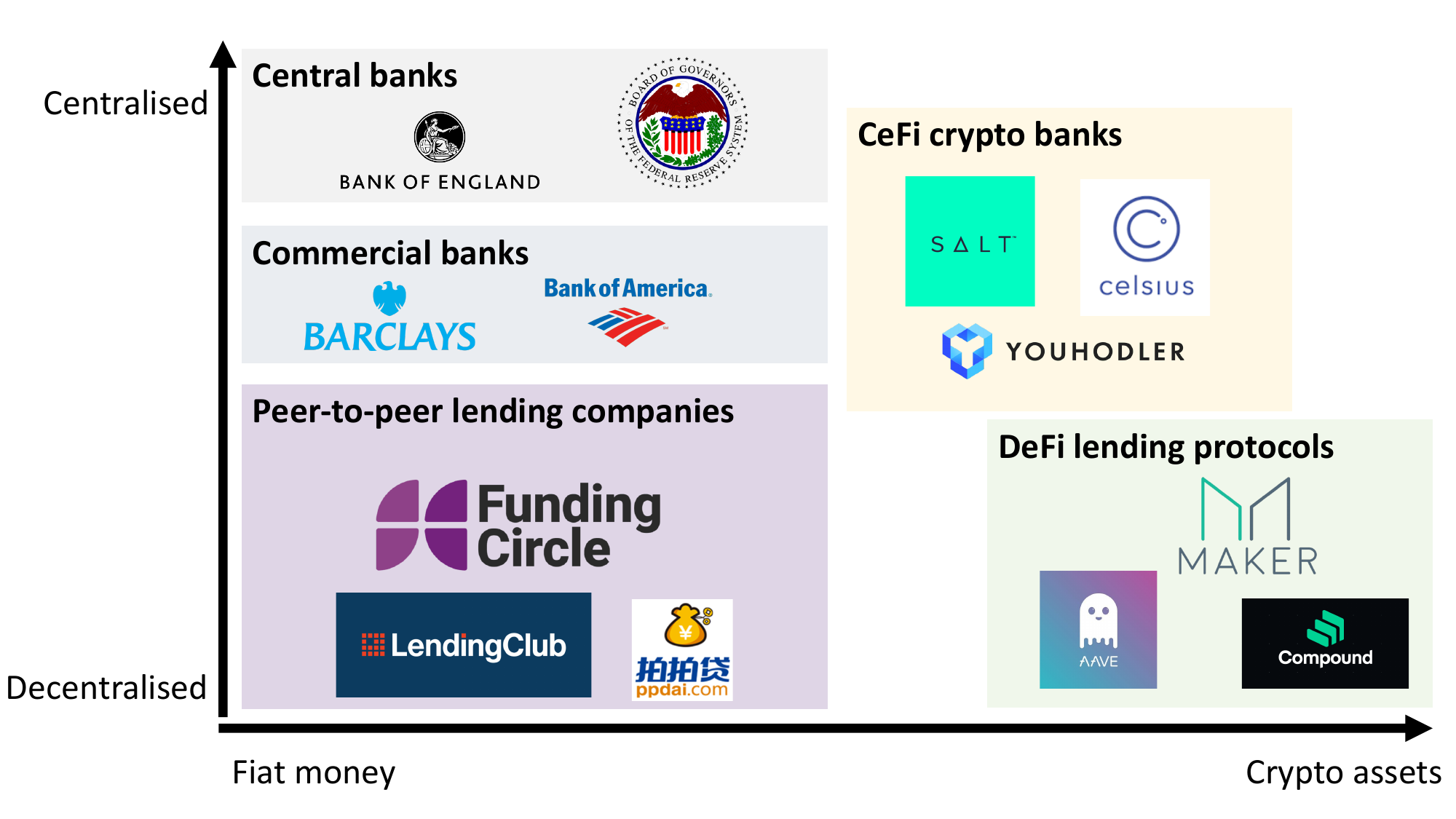}
    \caption{Taxonomy of lending markets}
    \label{fig:lendingmarkets}
\end{figure}

Centralised finance (CeFi) platforms for lending such as Salt, Celsius and YouHodler serve as bridges between the conventional monetary market and crypto-asset market. With those platforms, one can borrow fiat money directly (instead of fiat-pegged stablecoins), with their crypto holdings serving as collateral. Those platforms are operated by registered companies that act as counterparties to both their depositing and borrowing users. For that reason, those companies are also called crypto banks (see \autoref{fig:lendingmarkets}).

\section{Outlook of the lending market – the path to the Internet of Value}
\label{sec:outlook}

The emergence of the alternative finance sector focussed on lending has not led to the anticipated change and disruption to mainstream lending markets. The majority of these companies, reflected in the bottom-left part of \autoref{fig:lendingmarkets}, are yet to make a profit. While many have floated on international stock exchanges, their share prices have tumbled since their Initial Public Offerings \citep{Scott2019}. Their problem has not only been the high costs of funding versus the incumbent commercial banks. Their technology advantage and abandonment of legacy infrastructure have not sufficed to outplay the competition. Furthermore, consolidation is occurring in the traditional part of the sector \citep{Megaw2020}. Commercial banks are buying marketplace lenders who are in trouble and combining the latter’s technological advantage with their own lower funding costs. Hopefully, this can only be good news for borrowers.

The Internet of Value, with a full spectrum of DeFi lending enabled, will mean that borrowers and lenders can seamlessly find yield and maximise their economic returns with whatever tokens and assets they wish to hold, without needing to go through any centralised intermediaries.

Expectably, the current DeFi markets’ exceptional double-digit yields will fall as it is unlikely these yields will be sustainable. However, already they are beginning to attract interest from institutional players who have only invested in the traditional financial markets \citep{shen2021institutionsBitcoin}. Although there are many more esoteric notions with blockchain technology powering the Internet of Value, such as asset tokenisation, the concept of yield is much more familiar to financial services professionals. Therefore, traditional financial services companies may seek to incorporate DeFi into their offerings sooner than expected.

As things stand, the traditional money market underpinned by central banks will not disappear. DeFi still faces regulatory and technology maturity issues, and it remains to be seen what will happen when these systems scale. However, with increased interest in Central Bank Digital Currency (CBDC) development, the conventional money market and DeFi will converge in the foreseeable future.

\printbibliography
\end{document}